\documentclass{ptapap}
\usepackage{amsmath}
\usepackage{amssymb}
\usepackage{hyperref}

\DeclareRobustCommand{\ion}[2]{%
\relax\ifmmode
\ifx\testbx\f@series
{\mathbf{#1\,\mathsc{#2}}}\else
{\mathrm{#1\,\mathsc{#2}}}\fi
\else\textup{#1\,{\mdseries\textsc{#2}}}%
\fi}

\newenvironment{myfont}{\fontfamily{lmtt}\selectfont}{\par}
\DeclareTextFontCommand{\textmyfont}{\myfont}

\author{Swayamtrupta Panda}[CFT,LNA,CAMK]
\author{Alberto Rodr\'{\i}guez-Ardila}[LNA,INPE]
\affil[CFT]{Center for Theoretical Physics, Polish Academy of Sciences, Al. Lotnik\' ow 32/46, 02--668 Warsaw, Poland}
\affil[CAMK]{Nicolaus Copernicus Astronomical Center, Polish Academy of Sciences, Bartycka 18, 00--716 Warsaw, Poland}
\affil[LNA]{Laborat\'orio Nacional de Astrof\'isica - MCTIC, R. dos Estados Unidos, 154 - Na\c{c}\~oes, Itajub\'a - MG, 37504--364, Brazil}
\affil[INPE]{Divis\~ao de Astrof\'{\i}sica, INPE, Avenida dos Astronautas 1758, S\~ao Jos\'e dos Campos, 12227-010, SP, Brazil}

\title{A novel black hole mass scaling using coronal lines in active galaxies}

\begin{document}

\maketitle

\begin{abstract}
Using \textit{bona-fide} black hole (BH) mass estimates from reverberation mapping and the line ratio [\ion{Si}{vi}]~1.963$\mu$m/Br$\gamma_{\rm broad}$ as tracer of the Active Galactic Nuclei (AGN) ionizing continuum, we find a novel BH-mass scaling relation of the form log($M_{\rm BH}) = (6.40\pm 0.17) - (1.99\pm 0.37) \times$ log ([\ion{Si}{vi}]/Br$\gamma_{\rm broad})$, over the BH mass interval, $10^6 - 10^8$ M$_{\odot}$. The current sample consists of 21 Type-1 AGNs with the overall dispersion in our scaling relation at 0.47~dex, one that emulates the well-established M-$\sigma$ relation which shows a dispersion $\sim$0.44~dex. The new scaling offers an economic, and physically motivated alternative for BH estimate using single epoch spectra, avoiding large telescope time (reverberation mapping) or absolute flux calibration (the continuum luminosity method). With the advent of big surveys in the infrared in the near future, we aim to reduce the scatter in the relation by supplementing with more sources.

\end{abstract}

\section{Introduction}

The determination of black hole (BH) masses is one of the major focus in the studies of supermassive black holes and the active galaxies where they reside. Most BH mass estimates are based on correlations between the BH mass and the stellar bulge velocity dispersion, i.e., the M-$\sigma$ relation \citep[e.g.][]{ferrarese_merritt2000,gultekin+09}, or the AGN continuum luminosity, i.e., the mass-luminosity relation by which the optical, UV and X-ray luminosities are found to correlate with the size of the Broad Line Region (BLR) \citep[e.g.][and references therein]{koratkar_gaskell1991,kaspi_etal2005,landt+13}. While the use of the M-$\sigma$ relation requires the measurement of $\sigma$, it is not always easy to determine it, particularly in AGNs wherein the strong continuum from the nuclear region dilutes the stellar absorption lines. In order to overcome this difficulty, a number
of alternative scaling relations using emission lines such as [\ion{O}{iii}]~$\lambda$5007 to measure the mass of the bulge 
\citep[e.g.][]{nelson_whittle1996}, [\ion{O}{ii}]~$\lambda$3727 \citep[e.g.][]{salviander_etal2006}, H$\beta$ or H$\alpha$ \citep[e.g.][]{kaspi_etal2005,greene_ho2005} to infer on the BLR size, or [\ion{Fe}{ii}] in the near-infrared  \citep[e.g.][]{riffel_etal2013} to infer on the stellar $\sigma$, have been proposed. 
 
Coronal Lines (CLs) are emission lines with high ionisation potentials (IPs range between $\sim$50 eV up to few hundreds eV) which makes them excellent tracers of the ionising continuum. Although often fainter than the classical medium-ionisation lines used for photoionisation diagnosis, high angular resolution in nearby AGN has shown that CLs  particularly in the near-infrared (NIR) are among the most conspicuous features \citep[e.g.][]{marconi1994, muller-sanchez+11, rodriguez+17, gravity2020}. In this work, we highlight the dependence of the BH mass with one such CL, [\ion{Si}{vi}]~1.963$\mu$m (IP [\ion{Si}{vi}] = 167 eV) in the NIR\footnote{this CL is among the most common and brightest ones observed in spectra of AGNs \citep{rodriguez+11, lamperti+17}. We refer the readers to \citet{2020arXiv201000075R} for a broad overview of our presented results.} after normalising it to the nearest H\,{\sc i} broad line emission (in this case Br$\gamma$). A tight correlation between BH mass and the CL ratio [\ion{Si}{vi}]/Br$\gamma_{\rm broad}$  is observed.

\section{Coronal line diagnostic diagrams}

Objects in this work are selected by having BH masses determined by reverberation mapping and single epoch optical and/or near-IR spectra with accurate CL measurements.  The first criterion restricts the sample to Type-1 sources only. The second avoids variability issues. The final working sample of objects has 21 AGNs with well-defined [\ion{Si}{vi}]~1.963$\mu$m/Br$\gamma_{\rm broad}$ estimates  \citep[see Table 1 in][]{2020arXiv201000075R}.

\begin{figure}[!h]
    \centering
    \includegraphics[width=0.5\columnwidth]{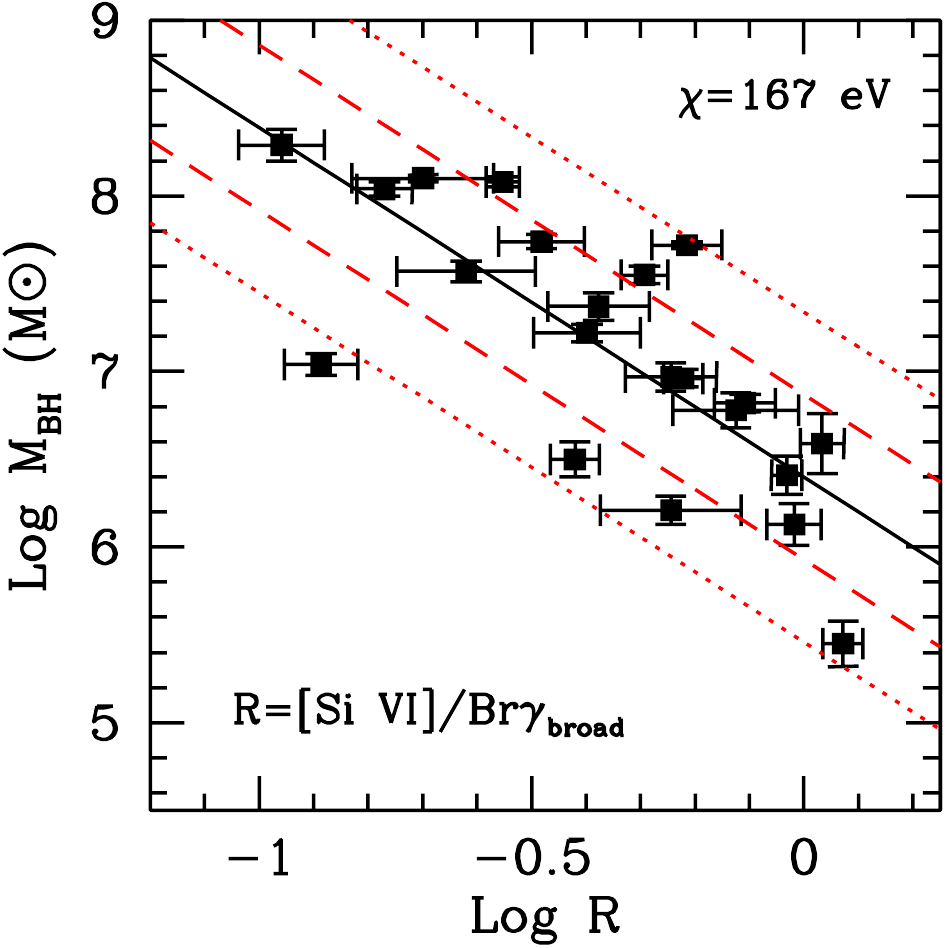}
    \caption{Observed [\ion{Si}{vi}]~1.963$\mu$m/Br$\gamma_{\rm broad}$ ratio versus black hole mass for the objects in our sample. The black line is the linear best-fit to the data and the red-dashed and -dotted lines show the 1$\sigma$ and 2$\sigma$ deviation, respectively. Figure courtesy: \citet{2020arXiv201000075R}.}
    \label{fig:my_label}
\end{figure}

Fig.~\ref{fig:my_label} presents a new diagnostic diagram in which the BH mass for the objects in our sample is plotted against the [\ion{Si}{vi}]~1.963$\mu$m/Br$\gamma_{\rm broad}$ ratio, which shows a clear trend with $M_{\rm BH}$ over three orders of magnitude in BH mass. A linear regression yields:
\begin{equation}
    \log M_{\rm BH} = (6.40\pm 0.17) - (1.99\pm 0.37) \times \log \left(\rm{\frac{[Si\,{\sc VI}]}{Br\gamma_{\rm broad}}}\right),
\end{equation}
and a 1$\sigma$ dispersion of 0.47 dex. The regression analysis follows the {\sc LtsFit} package\footnote{\href{http://www-astro.physics.ox.ac.uk/~mxc/software/\#lts}{http://www-astro.physics.ox.ac.uk/mxc/software/lts}} \citep{capellari+13}, which accounts for the errors in all variables. The Pearson correlation coefficient for the correlation is $r$ = -0.76, with a \textit{p}-value = 3.8$\times 10^{-5}$.

\section{Concluding remarks}
With a final compendium of 21 AGNs, the dispersion in BH mass in the proposed  calibration is 0.47 dex (1$\sigma$). In comparison,  a dispersion of  0.44 dex is inferred  from the $M-\sigma$ relation in 49 galactic bulges with direct dynamical BH mass estimate \citep{gultekin+09}. The intrinsic scatter in the mass - luminosity relations is in the 40\% range \citep{kaspi_etal2005}, mostly driven by differences in optical - UV continuum shape.

The present BH mass scaling relation is restricted to Type-1 AGN including narrow line Seyfert galaxies. The limitation is driven by the imposition of including  {\it bona-fide} BH masses only, and the need to normalise to broad \ion{H}{i} gas.  We are nonetheless examining possibilities to extend it to Type 2. The new scaling offers an economic and physically motivated alternative for BH estimate using single epoch spectra, avoiding  large telescope time (reverberation mapping) or absolute flux calibration (the continuum luminosity method). With James Webb Space Telescope and big surveys in the IR region, large samples of AGNs could be weighted using this approach. For a full account of the observed spectra for the sources, and detailed photoionisation modelling that confirms the origin of the observed scaling relation, we refer the readers to \citet{2020arXiv201000075R}.

\acknowledgements{The project was partially supported by the Polish Funding Agency National Science Centre, project 2017/26/\-A/ST9/\-00756 (MAESTRO  9), MNiSW grant DIR/WK/2018/12 and acknowledges partial support from CNPq Fellowship (164753/2020-6).}

\bibliographystyle{ptapap}
\bibliography{panda3}

\begin{thebibliography}{17}
\providecommand{\natexlab}[1]{#1}
\providecommand{\url}[1]{\texttt{#1}}
\providecommand{\urlprefix}{URL }
\providecommand{\eprint}[2][]{\url{#2}}

\bibitem[{{Cappellari} et~al.(2013)}]{capellari+13}
{Cappellari}, M., et~al., \emph{\mnras} \textbf{432}, 3, 1709 (2013)

\bibitem[{{Ferrarese} \& {Merritt}(2000)}]{ferrarese_merritt2000}
{Ferrarese}, L., {Merritt}, D., \emph{\apjl} \textbf{539}, 1, L9 (2000)

\bibitem[{{Gravity Collaboration} et~al.(2020)}]{gravity2020}
{Gravity Collaboration}, et~al., \emph{\aap} \textbf{643}, A154 (2020)

\bibitem[{{Greene} \& {Ho}(2005)}]{greene_ho2005}
{Greene}, J.~E., {Ho}, L.~C., \emph{\apj} \textbf{630}, 1, 122 (2005)

\bibitem[{{G{\"u}ltekin} et~al.(2009)}]{gultekin+09}
{G{\"u}ltekin}, K., et~al., \emph{\apj} \textbf{698}, 1, 198 (2009)

\bibitem[{{Kaspi} et~al.(2005)}]{kaspi_etal2005}
{Kaspi}, S., et~al., \emph{\apj} \textbf{629}, 1, 61 (2005)

\bibitem[{{Koratkar} \& {Gaskell}(1991)}]{koratkar_gaskell1991}
{Koratkar}, A.~P., {Gaskell}, C.~M., \emph{\apjl} \textbf{370}, L61 (1991)

\bibitem[{{Lamperti} et~al.(2017)}]{lamperti+17}
{Lamperti}, I., et~al., \emph{\mnras} \textbf{467}, 1, 540 (2017)

\bibitem[{{Landt} et~al.(2013)}]{landt+13}
{Landt}, H., et~al., \emph{\mnras} \textbf{432}, 1, 113 (2013)

\bibitem[{{Marconi} et~al.(1994){Marconi}, {Moorwood}, {Salvati}, \&
  {Oliva}}]{marconi1994}
{Marconi}, A., {Moorwood}, A.~F.~M., {Salvati}, M., {Oliva}, E., \emph{\aap}
  \textbf{291}, 18 (1994)

\bibitem[{{M{\"u}ller-S{\'a}nchez} et~al.(2011)}]{muller-sanchez+11}
{M{\"u}ller-S{\'a}nchez}, F., et~al., \emph{\apj} \textbf{739}, 2, 69 (2011)

\bibitem[{{Nelson} \& {Whittle}(1996)}]{nelson_whittle1996}
{Nelson}, C.~H., {Whittle}, M., \emph{\apj} \textbf{465}, 96 (1996)

\bibitem[{{Riffel} et~al.(2013)}]{riffel_etal2013}
{Riffel}, R.~A., et~al., \emph{\mnras} \textbf{429}, 3, 2587 (2013)

\bibitem[{{Rodr{\'\i}guez-Ardila} et~al.(2021){Rodr{\'\i}guez-Ardila}, {Panda},
  {Prieto}, \& {Marinello}}]{2020arXiv201000075R}
{Rodr{\'\i}guez-Ardila}, A., {Panda}, S., {Prieto}, A., {Marinello}, M.,
  \emph{\mnras} (in press) (2021)

\bibitem[{{Rodr{\'\i}guez-Ardila} et~al.(2011){Rodr{\'\i}guez-Ardila},
  {Prieto}, {Portilla}, \& {Tejeiro}}]{rodriguez+11}
{Rodr{\'\i}guez-Ardila}, A., {Prieto}, M.~A., {Portilla}, J.~G., {Tejeiro},
  J.~M., \emph{\apj} \textbf{743}, 2, 100 (2011)

\bibitem[{{Rodr{\'\i}guez-Ardila} et~al.(2017)}]{rodriguez+17}
{Rodr{\'\i}guez-Ardila}, A., et~al., \emph{\mnras} \textbf{470}, 3, 2845 (2017)

\bibitem[{{Salviander} et~al.(2006){Salviander}, {Shields}, {Gebhardt}, \&
  {Bonning}}]{salviander_etal2006}
{Salviander}, S., {Shields}, G.~A., {Gebhardt}, K., {Bonning}, E.~W.,
  \emph{\nar} \textbf{50}, 9-10, 803 (2006)

\end{thebibliography}

\end{document}